\documentclass[conference]{IEEEtran} 
\IEEEoverridecommandlockouts
\usepackage{cite}
\usepackage{url}
\usepackage{amsmath,amssymb,amsfonts}
\usepackage{algorithm}
\usepackage{algorithmic}
\usepackage{graphicx}
\usepackage{float}
\usepackage{subfigure}
\usepackage{subfig}
\graphicspath{{pics/},{pics/NUC/},{pics/BLER/}}
\newcommand{\figref}[1]{Fig.~\ref{#1}}
\usepackage{makecell}
\usepackage{threeparttable}
\usepackage{textcomp}
\usepackage{xcolor}
\usepackage{fancyhdr}
\def\BibTeX{{\rm B\kern-.05em{\sc i\kern-.025em b}\kern-.08em
    T\kern-.1667em\lower.7ex\hbox{E}\kern-.125emX}}
\begin{document}
\fancyhf{}
\pagestyle{empty}
\thispagestyle{empty}

\title{Capacity-based Spatial Modulation Constellation and Pre-scaling Design\\
}

\author{
    \IEEEauthorblockN{Xinghao Guo\IEEEauthorrefmark{1}, Hanjiang Hong\IEEEauthorrefmark{1}, Yin Xu\IEEEauthorrefmark{1}, Yi-yan Wu\IEEEauthorrefmark{2}, Dazhi He\IEEEauthorrefmark{1} and Wenjun Zhang\IEEEauthorrefmark{1}}
    \IEEEauthorblockA{\IEEEauthorrefmark{1} School of Electronic Information and Electrical Engineering, Shanghai Jiao Tong University,\\Shanghai 200240, China \\ Email: \{guoxinghao, honghj, xuyin, hedazhi, zhangwenjun\}@sjtu.edu.cn\\\IEEEauthorrefmark{2} Wireless Technology Research Department, Communications Research Centre, \\Ottawa ON K2K 2Y6, Canada \\ Email: yiyan.wu@ieee.org}
}

\maketitle

\begin{abstract}
     Spatial Modulation (SM) can utilize the index of the transmit antenna (TA) to transmit additional information. In this paper, to improve the performance of SM, a non-uniform constellation (NUC) and pre-scaling coefficients optimization design scheme is proposed. The bit-interleaved coded modulation (BICM) capacity calculation formula of SM system is firstly derived. The constellation and pre-scaling coefficients are optimized by maximizing the BICM capacity without channel state information (CSI) feedback. Optimization results are given for the multiple-input-single-output (MISO) system with Rayleigh channel. Simulation result shows the proposed scheme provides a meaningful performance gain compared to conventional SM system without CSI feedback. The proposed optimization design scheme can be a promising technology for future 6G to achieve high-efficiency. 
\end{abstract}


\section{Introduction} \label{i}
\thispagestyle{empty}
\IEEEPARstart{W}{ith} the massive growth of mobile devices and the widespread application of the Internet of Things (IoT), the sixth generation mobile communication (6G) is expected to achieve higher data transmission rates, spectral efficiency, and energy efficiency. There are various solutions in the physical layer to improve the efficiency and reliability, e.g., non-orthogonal multiple access (NOMA) \cite{NOMA}, massive multiple-input-multiple-output (MIMO) \cite{mMIMO}, reconfigurable intelligent surfaces (RIS) \cite{RIS,RIS-SM,RIS2}, etc.

Spatial modulation (SM), as a multi-antenna transmission technology, can achieve higher transmission capacity \cite{SM}. In SM, a single radio frequency (RF) chain is utilized in each slot to transmit symbols and the antenna index is encoded to transmit additional information bits. Compared to other MIMO transmission technologies, SM has the following benefits: high throughput \cite{STC}; reduced Inter-Antenna-Synchronization (IAS) and Inter Antenna Interference (IAI) \cite{IAS,IAI}; low power consumption; 
low detection complexity \cite{V-BLAST}. Due to the advantages mentioned above, SM is a promising new MIMO transmission technology and has become a hot spot in the MIMO research field.

In order to reduce the influence of fading channel on the system performance, much work has been done on the transmitter technology in SM. In \cite{yang}, Yang discusses a SM constellation optimization scheme which minimizes the bit error rate (BER). On the other hand, a scheme for designing SM non-uniform constellation (NUC) based on the system coded modulation average mutual information (CM-AMI) maximization criterion is studied in \cite{SM-NUC}. In fact, spatial shaping and symbol constellation shaping are discussed separately in the above work, so the transmission diversity cannot be improved. Besides, a joint design combining the selection of TAs and the transmit symbol constellation can achieve a higher gain. An optimal design algorithm combining phase rotation and power allocation is proposed in \cite{SM-CSI}, and the influence of channel estimation error is considered. Masouros proposed a phase rotation scheme to maximize the minimum Euclidean distance (MED) in the received constellation, eliminate the channel phase interference based on the channel state information (CSI), and combine the selection of TAs to pre-assign different phases \cite{SM-P}. Furthermore, the scheme that the transmitter and receiver simultaneously select the same randomly generated pre-scaling factor set is proposed to maximize the MED \cite{SM-CR}. The pre-scaling scheme is extended to the radar communication system \cite{Radar}. However, the CSI feedback plays an essential role in the above schemes that can mitigate the interference created by the channel. Designing robust pre-scaling strategies is an important topic to tackle, especially when the CSI feedback is unavailable. Therefore, this paper proposes a joint optimal design scheme of SM NUC and pre-scaling coefficients. The bit-interleaved coded modulation (BICM) capacity of SM system is derived to be maximized. Compared to conventional SM, the simulation demonstrates that the proposed scheme has up to 1.1 dB and 2.1 dB performance gain in the case of two different TA numbers.

The rest of this paper is organized as follows: Section \ref{ii} briefly introduces the SM system model and the principle of pre-scaling. Section \ref{iii} proposes the capacity-based joint optimization scheme to obtain the SM constellations and pre-scaling coefficients. Section \ref{iv} presents the optimization results and error performance analysis, and Section \ref{v} concludes the paper.

\section{Spatial modulation with pre-scaling} \label{ii}

\subsection{System Model}
\begin{figure}[htbp]
    \centerline{\includegraphics[width=0.5\textwidth]{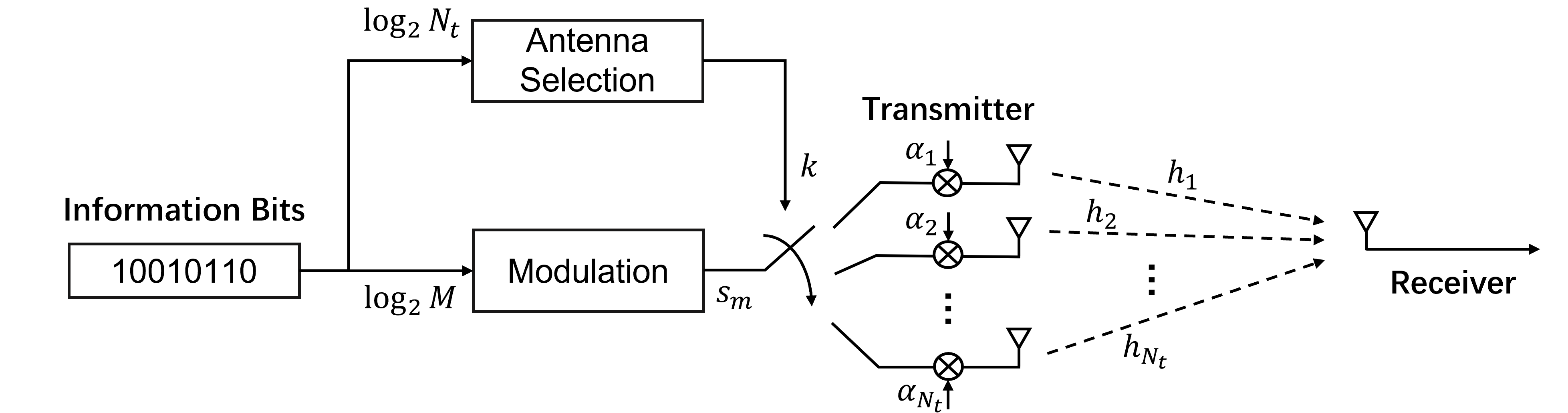}}
    \caption{Block diagram of SM transmitter with pre-scaling.}
    \label{model}
\end{figure}
Consider a multiple-input-single-output (MISO) system where the transmitter is equipped with $N_{t}$ antennas, as illustrated in \figref{model}. At the transmitter, the information bits are divided into two groups, one for mapping symbols and one for selecting TAs. Let $\mathbf{s}$ represent one SM transmit vector, so we have $\mathbf{s}=\mathbf{e}_{k}s_{m}^{k}$, in which $\mathbf{e}_{k}$ is a $N_{t}$-dimensional vector with $k$-th element is one and the rest are zeros, and $s_{m}^{k}$ is the $m$-th complex symbol chosen from the $M$-order constellation. Further, let $\chi_{k}=\{s_{1}^{k},s_{2}^{k},...,s_{M}^{k}\}$ represent the constellation used for transmission when the $k$-th TA is selected. So the transmit vectors set $\chi=\{\mathbf{e}_{k}s_{m}^{k}|s_{m}^{k} \in \chi_{k}, k=1,2,...,N_{t}\}$ can be expressed as \eqref{vector}, and each element represents a transmit vector.

\begin{small}   
\begin{equation}
\label{vector}
\setlength{\arraycolsep}{2.0pt}
\chi=
\begin{Bmatrix}
\!\!\begin{bmatrix} s_{1}^{1} \\ 0 \\ \vdots \\ 0 \end{bmatrix}\!\!,& 
\!\!\cdots\!\!&, 
\!\!\begin{bmatrix} s_{M}^{1} \\ 0 \\ \vdots \\ 0  \end{bmatrix}\!\!,& 
\!\!\begin{bmatrix} 0 \\ s_{1}^{2} \\ \vdots \\ 0 \end{bmatrix}\!\!,& 
\!\!\cdots\!\!&, 
\!\!\begin{bmatrix} 0 \\ s_{M}^{2} \\ \vdots \\ 0 \end{bmatrix}\!\!,& 
\!\!\cdots\!\!&, 
\!\!\begin{bmatrix} 0 \\ 0 \\ \vdots \\ s_{1}^{N_t} \end{bmatrix}\!\!,& 
\!\!\cdots\!\!&, 
\!\!\begin{bmatrix} 0 \\ 0 \\ \vdots \\ s_{M}^{N_t} \end{bmatrix}\!\!
\end{Bmatrix}
\end{equation}
\end{small}


Therefore, given the number of TAs $N_{t}$ and the modulation order of the constellation $M$, the spectral efficiency (SE) of SM is $\log_{2}{N_t}+\log_{2}{M}$ bits per channel use.
For fair comparison, constellation normalization principle is adopted in this paper, which satisfies:
\begin{equation}
    \label{nom0}
    P=\frac{1}{M} \sum_{m=1}^{M}\left|s_{m}^{k}\right|^{2} =1,\quad k=1,2,...,N_{t}
\end{equation}
where $P$ is the average power of the transmit vector, and $\left|x\right|$ denotes the amplitude of a complex number $x$.

At the receiver, the received signal is given by: 
\begin{equation}
    \label{yx}
    \mathbf{y}=\mathbf{H}\mathbf{s}+\mathbf{z} 
\end{equation}
where $\mathbf{H}= [h_{1},h_{2},\cdots,h_{N_{t}}]$ is the MISO channel vector, $h_{k}$ denotes the complex channel coefficient from the $k$-th TA to the receiver, $\mathbf{s} \in \chi$ is the transmit vector without pre-scaling, and $\mathbf{z} \sim \mathcal{C N}\left(0, \sigma^2 \mathbf{I}\right)$ is the additive white Gaussian noise (AWGN). For detection, under the assumption of complete channel knowledge, the principle with the optimal maximum likelihood (ML) detection can be expressed as \cite{ML}
\begin{equation}
    \label{ML}
    \hat{\mathbf{s}}=\underset{\mathbf{s}^{\prime} \in \chi}{\arg \min }\|\mathbf{y}-\mathbf{H} \mathbf{s}^{\prime}\|
\end{equation}
where $\|\mathbf{x}\|$ denotes the norm of vector $\mathbf{x}$.

\subsection{Transmit Pre-scaling}
In order to improve the diversity of SM, the pre-scaling approach is applied to the transmitter, as shown in \figref{model}. The signal fed to each TA is multiplied by a complex coefficient $\alpha_k$, and \eqref{nom0} can be rewritten as 
\begin{equation}
    \label{nom}
    \frac{1}{MN_{t}}\sum_{k=1}^{N_{t}} \sum_{m=1}^{M}\left|\alpha_ks_{m}^{k}\right|^{2} =1
\end{equation}

At the receiver, the received signal can be rewritten as 
\begin{equation}
    \label{yax}
    \mathbf{y}=\mathbf{H}\mathbf{A}\mathbf{s}+\mathbf{z} 
\end{equation}
where $\mathbf{A}=\operatorname{diag}(\mathbf{a}) \in \mathbb{C}^{N_t \times N_t}$ is the pre-scaling diagonal matrix with its diagonal elements taken from the vector $\mathbf{a}= [\alpha_{1},\alpha_{2},\cdots,\alpha_{N_{t}}]$. 

The existing SM pre-scaling schemes aim to maximize the MED between the two received signals $\mathbf{y}_i$ and $\mathbf{y}_j$. The optimal matrix $\mathbf{A}$ is obtained by solving the following optimization problem:
\begin{equation}
    \label{opt}
    \begin{aligned}
    &\mathbf{A}^*= \arg \max_{\mathbf{A}} \min_{i,j} \left\|\mathbf{y}_i-\mathbf{y}_j\right\|^2, i \neq j \\
    &\begin{array}{r@{\quad}r@{}l@{\quad}l}
    \text{s.t.} &\frac{1}{MN_{t}}\sum_{k=1}^{N_{t}} \sum_{m=1}^{M}\left|\alpha_ks_{m}^{k}\right|^{2} =1 \\
    \end{array}
    \end{aligned}
\end{equation}
In \cite{SM-P}, the pre-scaling coefficient $\alpha_k$ takes the following value:
\begin{equation}
    \label{sm-p}
    \alpha_k  =e^{j (\theta_k-\vartheta_k)} 
\end{equation}
where $\vartheta_k$ is the phase of the $k$-th channel coefficient, i.e.,  $h_{k}=\left|h_{k}\right|e^{j \vartheta_k}$, and $\theta_k$ is the $k$-th angle taken from an equally spaced arrangement within $[0,2\pi/M)$, which satisfies:
\begin{equation}
    \label{sm-p-angle}
    \theta_k=\frac{2\pi(k-1)}{MN_t}, \quad k=1,2,...N_t 
\end{equation}
In this scheme, when $\mathbf{s}_{i=(k-1)N_t+m}=\mathbf{e}_{k}s_{m}^{k}$ is transmitted, the received signal is:
\begin{equation}
    \label{sm-p-y}
    \begin{aligned}
    \mathbf{y}_i &= \mathbf{H}\mathbf{A}\mathbf{s}_i+\mathbf{z} \\
                &= h_{k}\alpha_ks_{m}^{k}+\mathbf{z} \\
                &= \left|h_{k}\right|e^{j \theta_k}s_{m}^{k}+\mathbf{z}
    \end{aligned}
\end{equation}
According to \eqref{sm-p-y}, the phases of the signal in the receive constellation satisfy the equidistant distribution, so it satisfies the MED maximization in the sense of phase. 

However, existing pre-scaling schemes are designed based on CSI feedback. Moreover, modern communication systems are mainly BICM systems, consisting of the forward error correction (FEC) encoder, bit interleaver, and constellation modulator \cite{BICM}. BICM capacity can represent the performance of constellations in a given channel. Capacity-based constellation with pre-scaling design is needed as an alternative without CSI feedback.

\section{Proposed Constellation Design with Pre-scaling}  \label{iii}
To fill the aforementioned gap, we devise capacity-based SM NUC and corresponding pre-scaling coefficients design method. The proposed scheme optimizes both the symbol coordinates of SM NUC and pre-scaling coefficients with the help of particle swarm optimization (PSO) algorithm.

\subsection{BICM Capacity of the SM System}\label{AA}
According to the calculation formula of the BICM average mutual information (BICM-AMI) derived in \cite{BICM}, the BICM-AMI of the above SM system can be derived as follows:
\begin{equation}
    \label{BICM-AMI}
    C=\log_{2}{MN_t}-\!\!\!\!\!\!\sum_{i=1}^{\log_{2}{MN_t}}\!\!\!\!\!\! E_{b, \mathbf{y,H}}\left[\log_2 \frac{\sum_{\mathbf{x} \in \chi} p(\mathbf{y} \mid \mathbf{HAx})}{\sum_{\mathbf{x} \in \chi_b^i} p(\mathbf{y} \mid \mathbf{HAx})}\right]
\end{equation}
where $E(\cdot)$ represents expectation calculation, $\chi$ is the transmit vector set given by \eqref{vector}, $\chi_b^i(b\in\{0,1\})$ denotes the transmit vectors set with the $i$-th bit being $b$, and $p(\cdot|\cdot)$ represents the conditional transition probability density function (PDF). According to properties of Gaussian distribution, there is
\begin{equation}
    \label{probability}
    p(\mathbf{y} \mid \mathbf{HAx})=\frac{1}{2 \pi \sigma^2} e^{-\frac{\|\mathbf{y}-\mathbf{HAx}\|^2}{2 \sigma^2}}
\end{equation}
where $\sigma^2$ denotes the AWGN variance on each dimension of $\mathbf{HAx}$.

\subsection{Optimization Algorithm with PSO}
According to the calculation formula of BICM-AMI \eqref{BICM-AMI} and the power normalization constraint \eqref{nom}, the optimization problem can be expressed as
\begin{equation}
    \label{opt-C}
    \begin{aligned}
    &\max_{\chi,\mathbf{A}}\quad C(\chi,\mathbf{A}) \\
    &\begin{array}{r@{\quad}r@{}l@{\quad}l}
    \text{s.t.} &\frac{1}{MN_{t}}\sum_{k=1}^{N_{t}} \sum_{m=1}^{M}\left|\alpha_ks_{m}^{k}\right|^{2} =1 \\
    \end{array}
    \end{aligned}
\end{equation}

To reduce optimization complexity without losing generality, assume that each TA uses the same constellation, i.e., $s_m^i=s_m^j=s_m, 1\leq i \neq j\leq N_t$, and that the constellation satisfies quadrantal symmetry, i.e., for $1\leq m\leq M/4$, satisfy:
\begin{equation}
    \label{symmetry}
    s_{m+M/4}=-s_m^*,\;\;s_{m+M/2}=s_m^*,\;\;s_{m+3M/4}=-s_m
\end{equation}

\begin{figure}[htbp]
    \centerline{\includegraphics[width=0.35\textwidth]{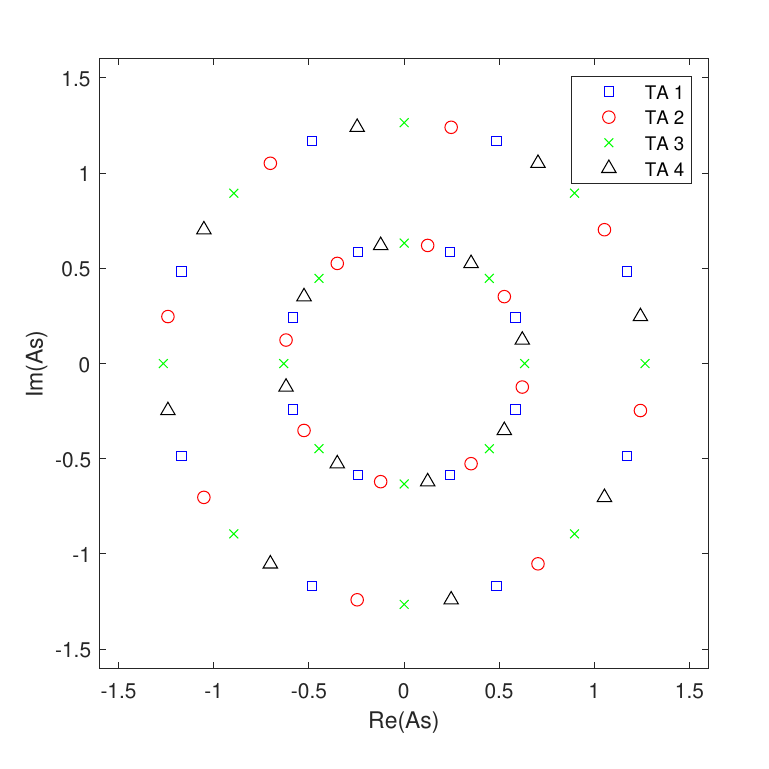}}
    \caption{Initial constellation with pre-scaling when $M=16$ and $N_t=4$.}
    \label{initial}
\end{figure}

Before performing the optimization algorithm, the initial values of the variables to be optimized need to be set first. Amplitude and phase-shift keying (APSK) is adopted to design the initial constellation. For APSK with modulation order $M=2^{m}$, constellation symbols are scattered on $2^{m/2-1}$ concentric rings with uniformly increasing radius and each ring contains $2^{m/2+1}$ uniform-distributed constellation symbols. Inspired by \cite{SM-P}, the initial values of the pre-scaling coefficients are as follows: 
\begin{equation}
    \label{pre-scaling}
    \alpha_k  =e^{j \frac{\pi(k-1)}{2^{m/2}N_t}}, \quad k=1,2,...N_t 
\end{equation}
For example, in the case of $M=16$ and $N_t=4$, the initial constellation with pre-scaling is shown in \figref{initial}, where the different marks represent the symbols eventually transmitted by the corresponding TA.

With the above preparation work, the specific optimization process is elaborated in \textbf{Algorithm \ref{algorithm1}}. According to the initial variable, the signal-to-noise ratio (SNR) threshold is obtained by simulation first. Then BICM-AMI is calculated and the optimization variables are updated with the PSO algorithm. Repeat the above process and iterate until the loop stop condition is met.
\begin{algorithm}[H]
	\setcounter{algorithm}{0}       
    \caption{NUC and pre-scaling coefficients optimization based on PSO.}     
	\label{algorithm1} 
	\renewcommand{\algorithmicrequire}{\textbf{Initialization:}}
	\renewcommand{\algorithmicensure}{\textbf{Output:}}
	\begin{algorithmic}[1]
		\REQUIRE initial constellation, initial pre-scaling coefficients, error threshold $\xi$.
		\STATE Simulate to get the waterfall SNR threshold $SNR_0$ with a given FEC code;
		\STATE $i \gets 0$ 
		\LOOP 
		    \STATE $i \gets i+1$;
		    \STATE Substitute the NUC, pre-scaling coefficients and $SNR_{i-1}$ into formula \eqref{BICM-AMI} to calculate the current BICM-AMI $C$ by Monte Carol method;
		    \STATE Update NUC and pre-scaling coefficients with PSO based on $C$, and satisfies the constraints \eqref{nom} and \eqref{symmetry};
		    \STATE Simulate to get the new waterfall SNR threshold $SNR_i$;
		    \IF{$SNR_{i-1}-SNR_{i}\leq \xi$}
		        \STATE \textbf{break};
		    \ENDIF
		\ENDLOOP
		\ENSURE the final NUC and pre-scaling coefficients. 
	\end{algorithmic} 
\end{algorithm}

\section{Performance Evaluation}  \label{iv}
\begin{figure}[htbp]
    \centering
    \begin{minipage}[b]{0.32\linewidth}
        \centering
        \includegraphics[width=\linewidth]{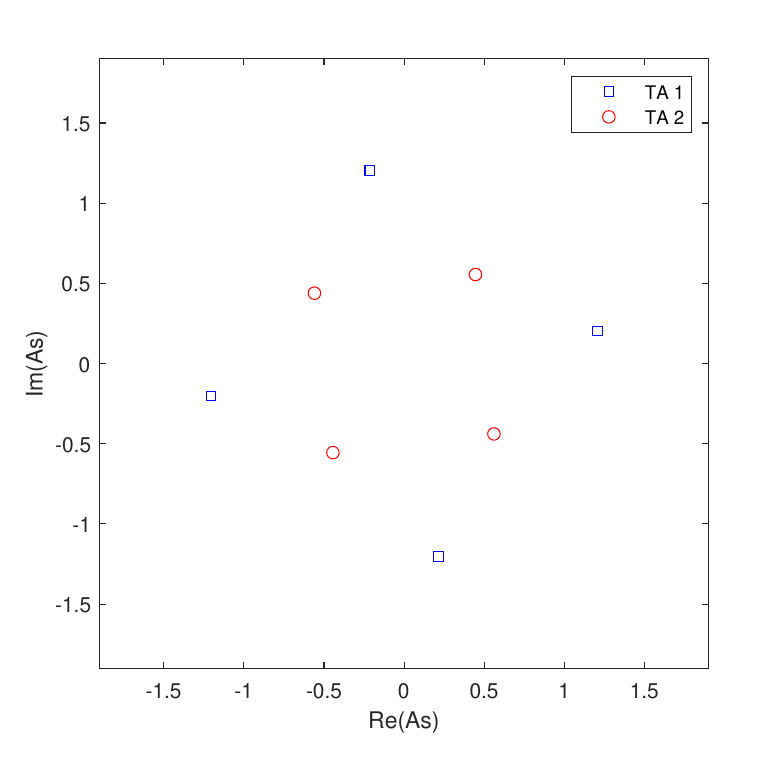}
        \caption*{(a)}
        \label{fig:2MCS0}
    \end{minipage}
    \begin{minipage}[b]{0.32\linewidth}
        \centering
        \includegraphics[width=\linewidth]{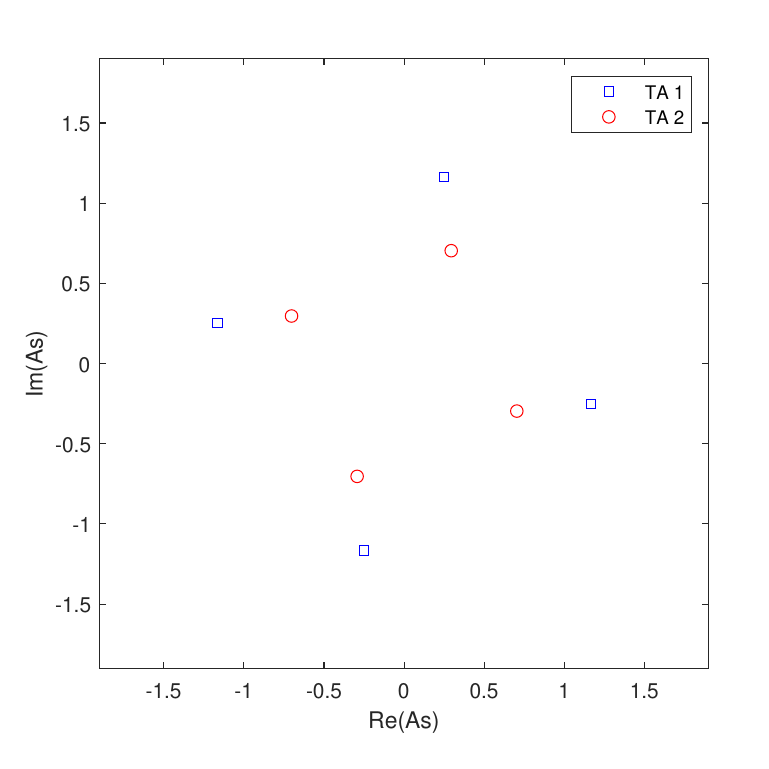}
        \caption*{(b)}
        \label{fig:2MCS4}
    \end{minipage}
    \begin{minipage}[b]{0.32\linewidth}
        \centering
        \includegraphics[width=\linewidth]{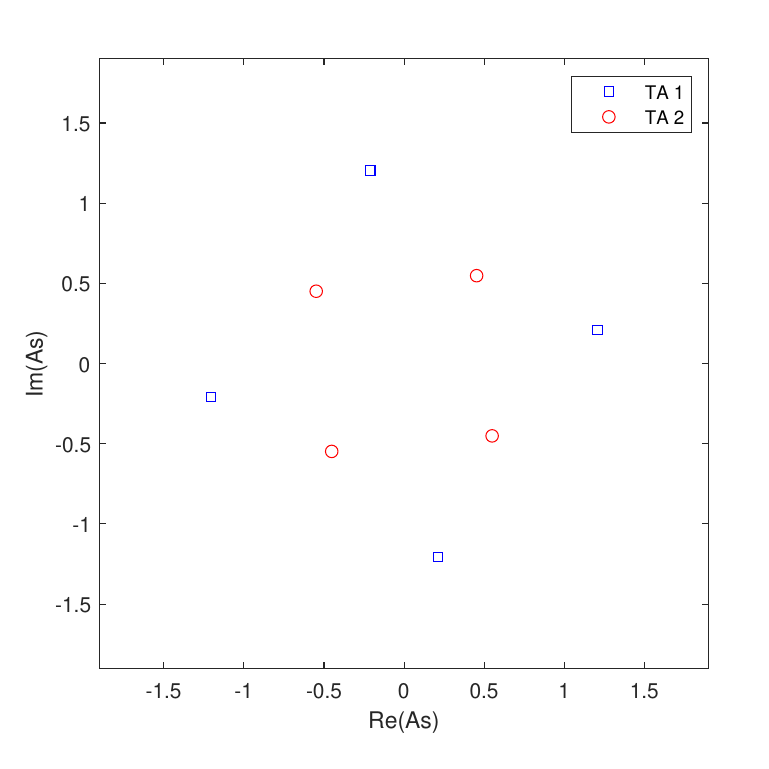}
        \caption*{(c)}
        \label{fig:2MCS9}
    \end{minipage}
    \medskip
    \begin{minipage}[b]{0.32\linewidth}
        \centering
        \includegraphics[width=\linewidth]{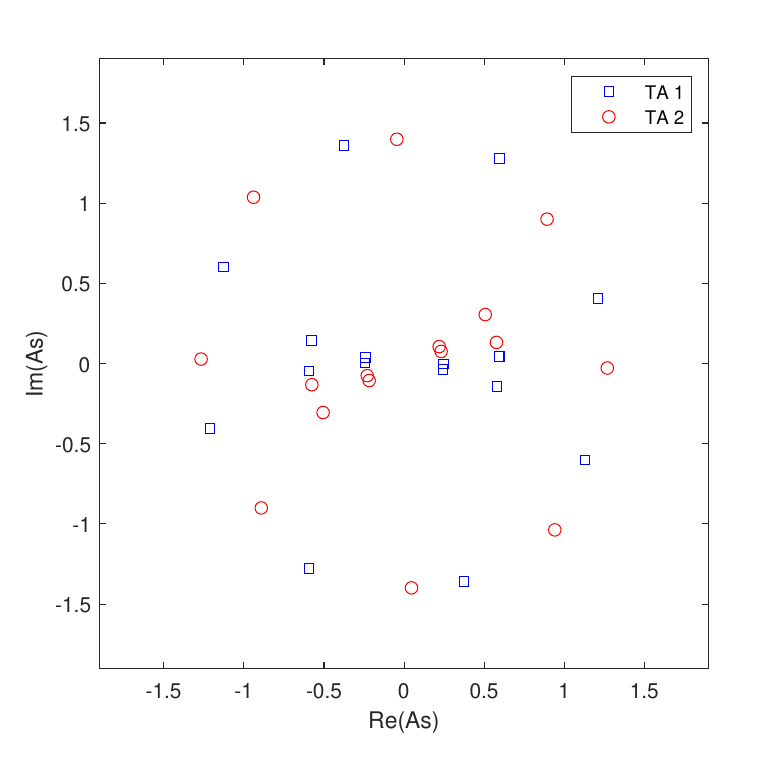}
        \caption*{(d)}
        \label{fig:2MCS10}
    \end{minipage}
    \begin{minipage}[b]{0.32\linewidth}
        \centering
        \includegraphics[width=\linewidth]{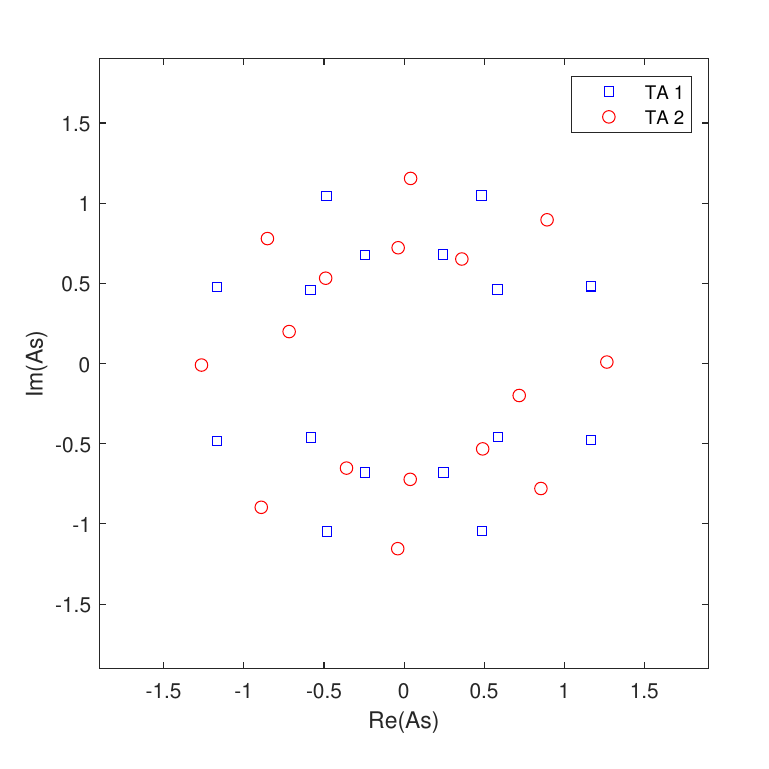}
        \caption*{(e)}
        \label{fig:2MCS13}
    \end{minipage}
    \begin{minipage}[b]{0.32\linewidth}
        \centering
        \includegraphics[width=\linewidth]{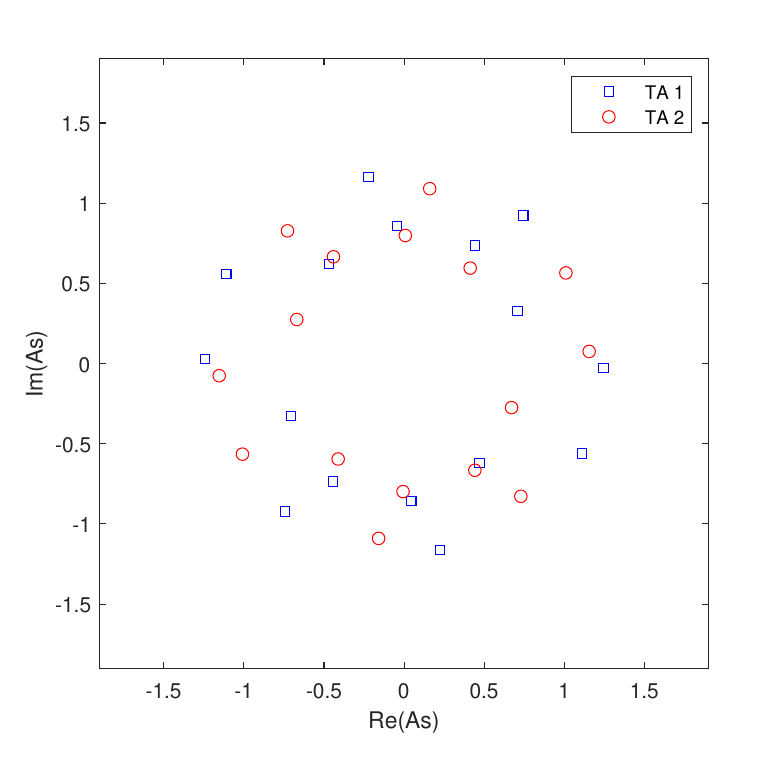}
        \caption*{(f)}
        \label{fig:2MCS16}
    \end{minipage}
    \medskip
    \begin{minipage}[b]{0.32\linewidth}
        \centering
        \includegraphics[width=\linewidth]{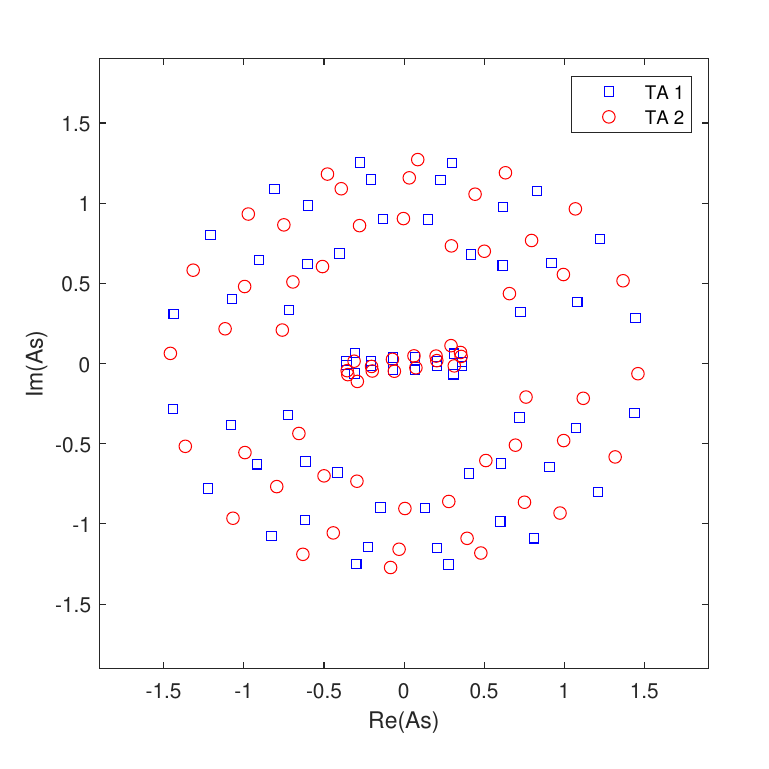}
        \caption*{(g)}
        \label{fig:2MCS17}
    \end{minipage}
    \begin{minipage}[b]{0.32\linewidth}
        \centering
        \includegraphics[width=\linewidth]{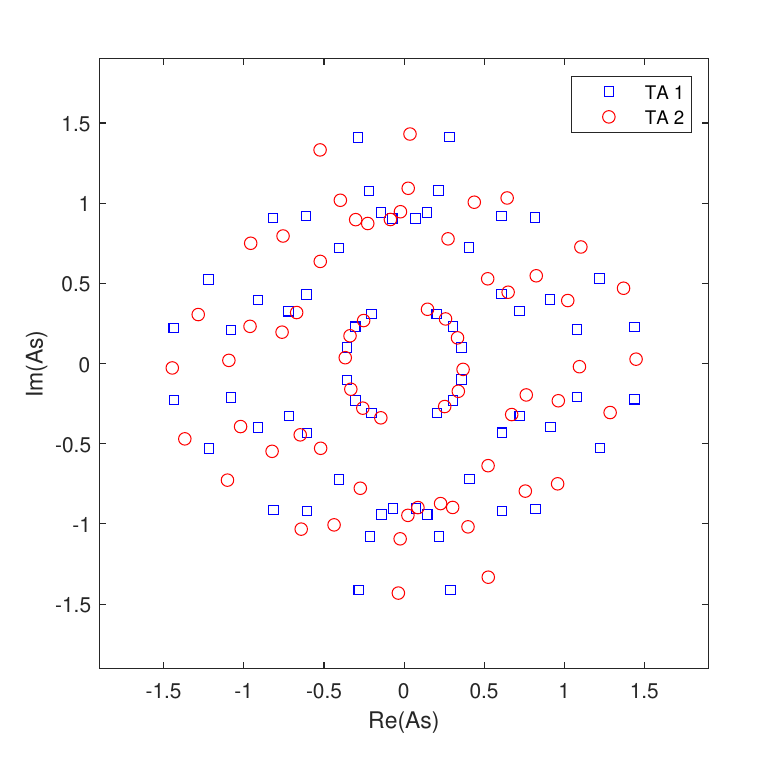}
        \caption*{(h)}
        \label{fig:2MCS22}
    \end{minipage}
    \begin{minipage}[b]{0.32\linewidth}
        \centering
        \includegraphics[width=\linewidth]{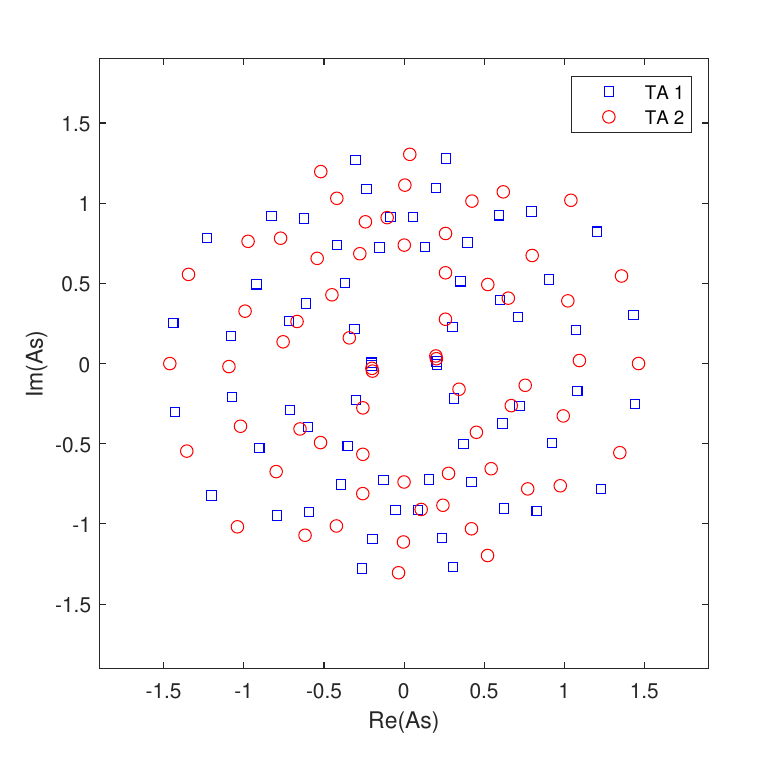}
        \caption*{(i)}
        \label{fig:2MCS28}
    \end{minipage}
    \caption{The NUCs with pre-scaling for $N_t=2$ and MCS = (a) 0,(b) 4,(c) 9,(d) 10,(e) 13,(f) 16,(g) 17,(h) 22,(i) 28.}
    \label{fig:2MCS}
\end{figure}
This section presents the simulation results of the optimization scheme proposed in \ref{iii}, and evaluate the SM system performance with the optimized NUC and pre-scaling coefficients. A Rayleigh channel is used whose CSI is completely unknown at the transmitter. The simulation bandwidth is 5.0 MHz, the available bandwidth (ABW) is 4.5 MHz, the subcarrier spacing $\Delta f=15$ kHz. The 5G New Radio (NR) low density parity check (LDPC) code is used. We use the modulation and coding schemes (MCSs) defined in \cite{3GPP} (Table 5.1.3.1-1). Transport block size (TBS) is then determined according to \cite{3GPP} (Section 5.1.3.2). 4$\times1$ and 2$\times$1 MISO systems are explored based the Max-Log-maximum a posteriori (Max-Log-MAP) demapping algorithm with ideal channel estimation.
\begin{figure}[htbp]
    \centering
    \begin{minipage}[b]{0.32\linewidth}
        \centering
        \includegraphics[width=\linewidth]{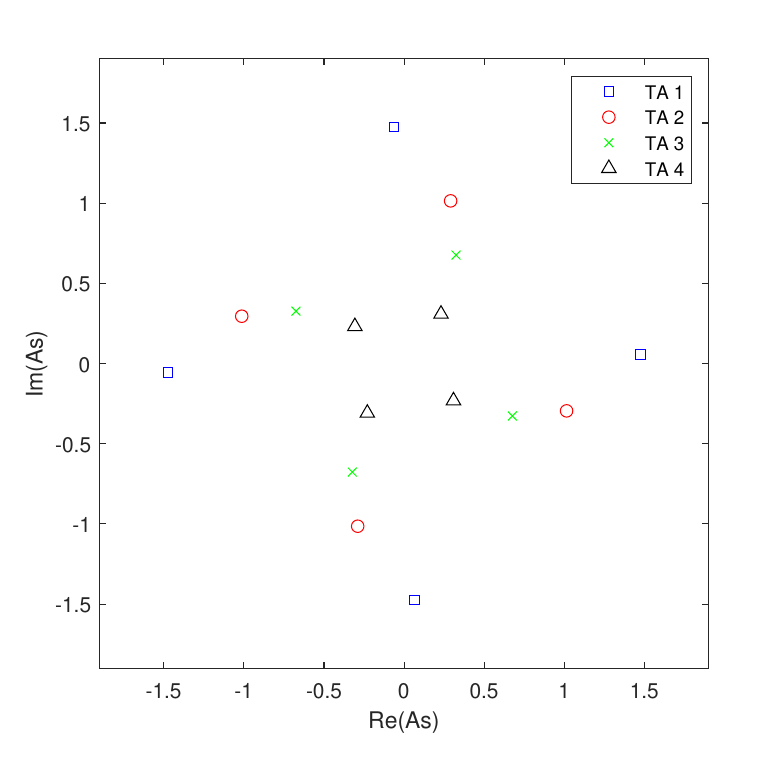}
        \caption*{(a)}
        \label{fig:4MCS0}
    \end{minipage}
    \begin{minipage}[b]{0.32\linewidth}
        \centering
        \includegraphics[width=\linewidth]{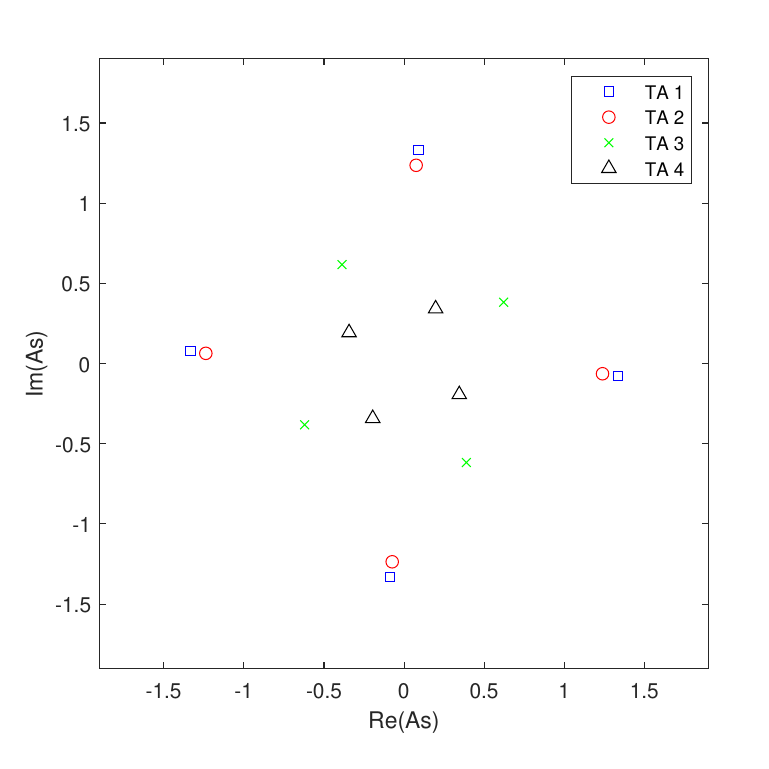}
        \caption*{(b)}
        \label{fig:4MCS4}
    \end{minipage}
    \begin{minipage}[b]{0.32\linewidth}
        \centering
        \includegraphics[width=\linewidth]{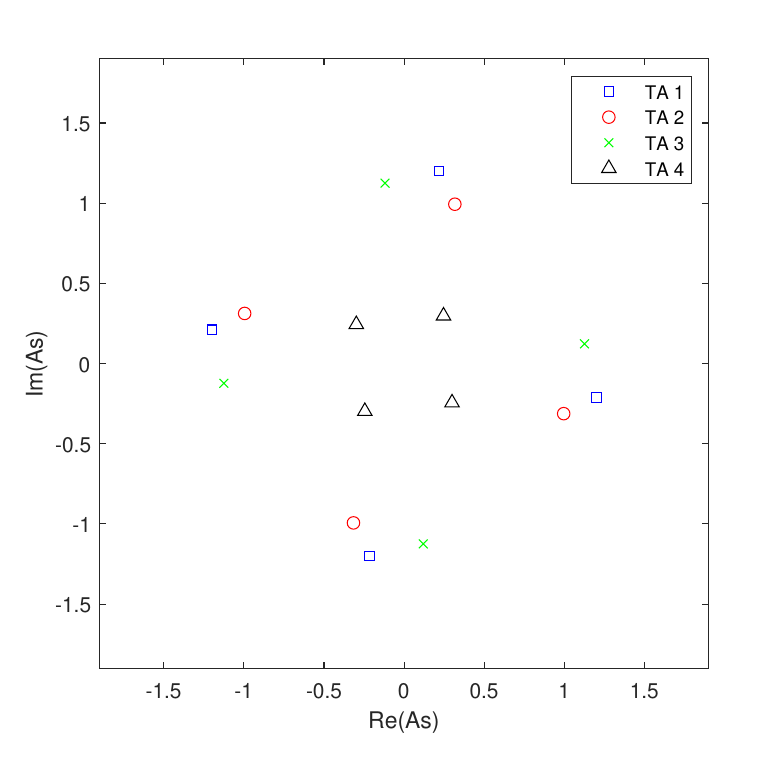}
        \caption*{(c)}
        \label{fig:4MCS9}
    \end{minipage}
    \medskip
    \begin{minipage}[b]{0.32\linewidth}
        \centering
        \includegraphics[width=\linewidth]{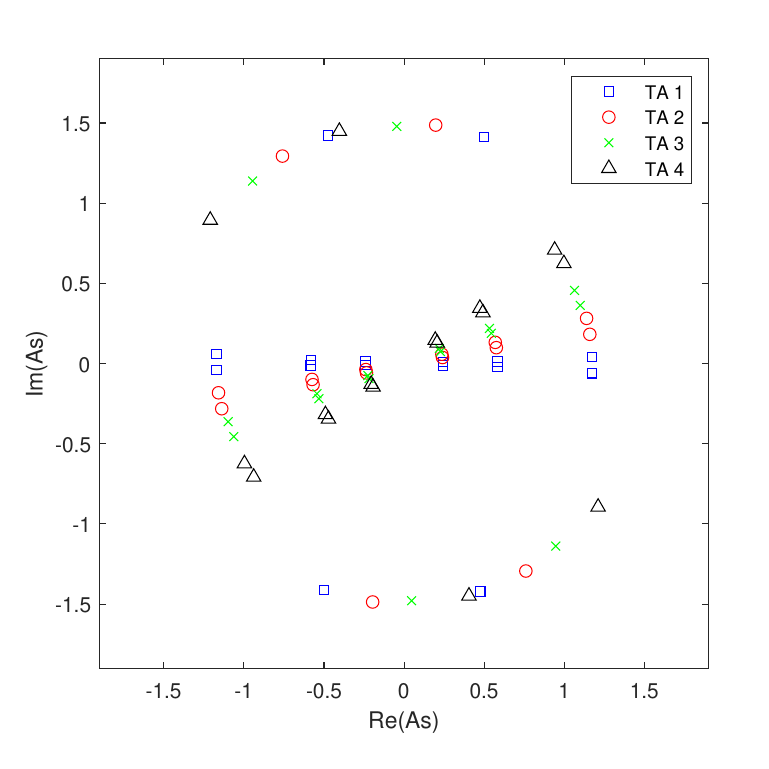}
        \caption*{(d)}
        \label{fig:4MCS10}
    \end{minipage}
    \begin{minipage}[b]{0.32\linewidth}
        \centering
        \includegraphics[width=\linewidth]{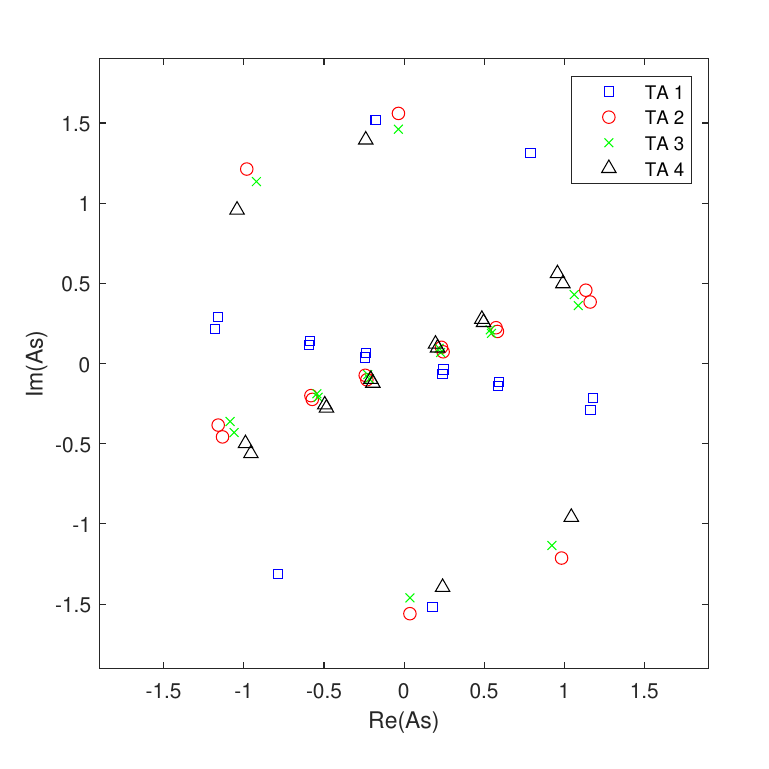}
        \caption*{(e)}
        \label{fig:4MCS13}
    \end{minipage}
    \begin{minipage}[b]{0.32\linewidth}
        \centering
        \includegraphics[width=\linewidth]{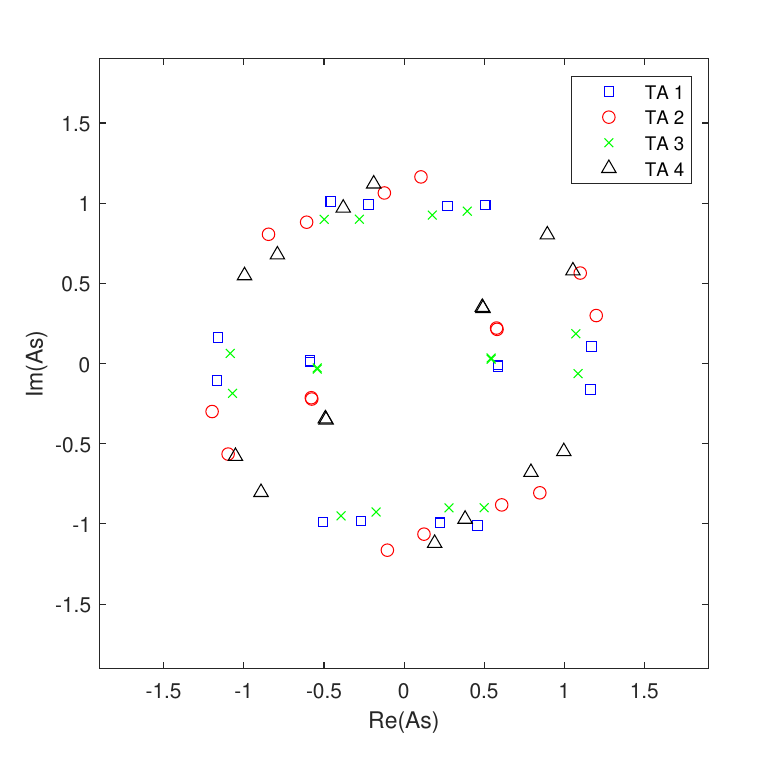}
        \caption*{(f)}
        \label{fig:4MCS16}
    \end{minipage}
    \medskip
    \begin{minipage}[b]{0.32\linewidth}
        \centering
        \includegraphics[width=\linewidth]{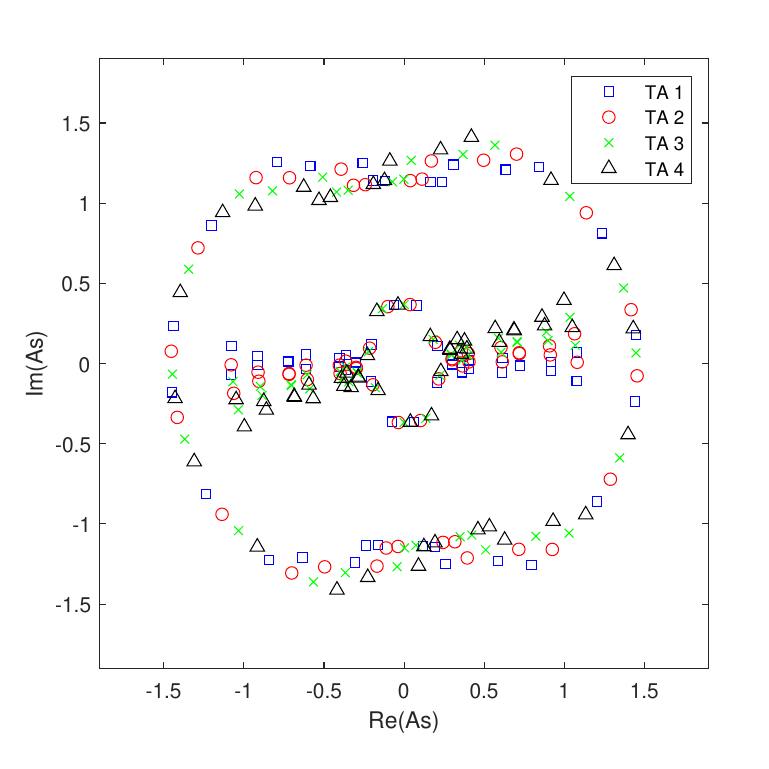}
        \caption*{(g)}
        \label{fig:4MCS17}
    \end{minipage}
    \begin{minipage}[b]{0.32\linewidth}
        \centering
        \includegraphics[width=\linewidth]{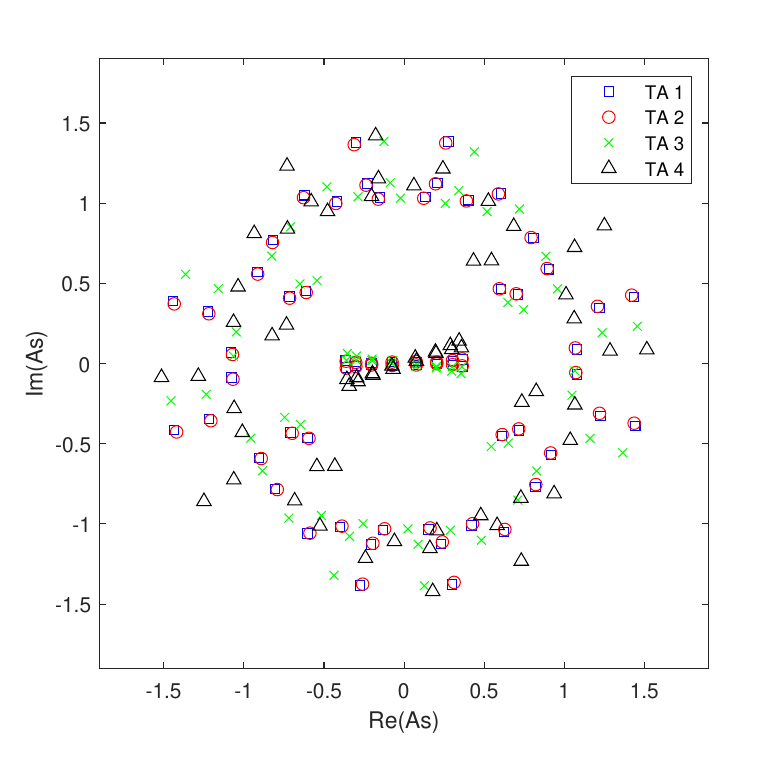}
        \caption*{(h)}
        \label{fig:4MCS22}
    \end{minipage}
    \begin{minipage}[b]{0.32\linewidth}
        \centering
        \includegraphics[width=\linewidth]{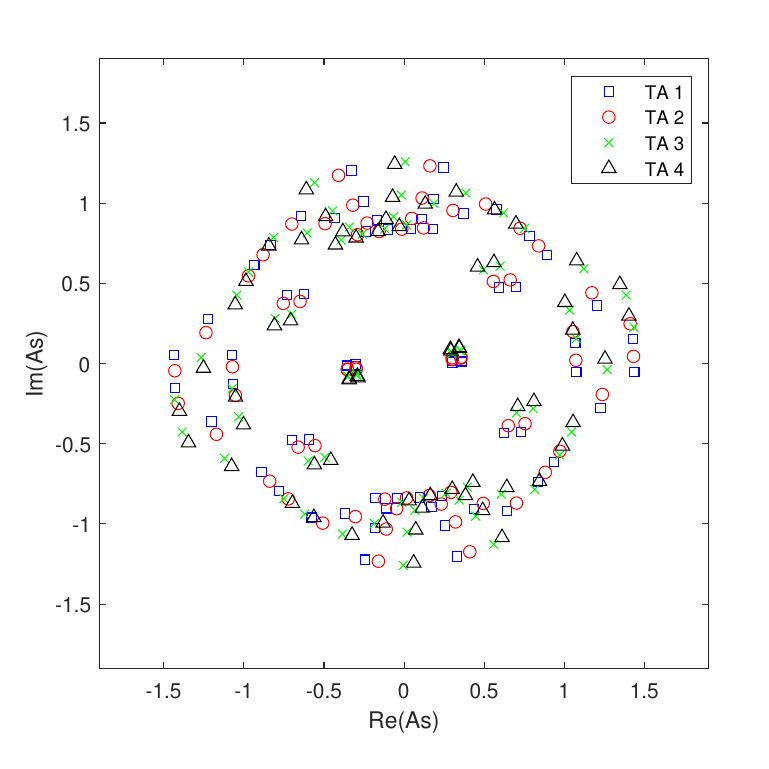}
        \caption*{(i)}
        \label{fig:4MCS28}
    \end{minipage}
    \caption{The NUCs with pre-scaling for $N_t=4$ and MCS = (a) 0,(b) 4,(c) 9,(d) 10,(e) 13,(f) 16,(g) 17,(h) 22,(i) 28.}
    \label{fig:4MCS}
\end{figure}

We select nine MCSs, namely 0,4,9,10,13,16,17,22,28, to test the proposed  optimization algorithm. For 2$\times$1 MISO system, the optimal NUCs with corresponding pre-scaling  coefficients are presented in \figref{fig:2MCS}, and \figref{fig:4MCS} shows the optimization results for 4$\times$1 MISO system.

\begin{figure}[htbp]
    \centerline{\includegraphics[width=0.4\textwidth]{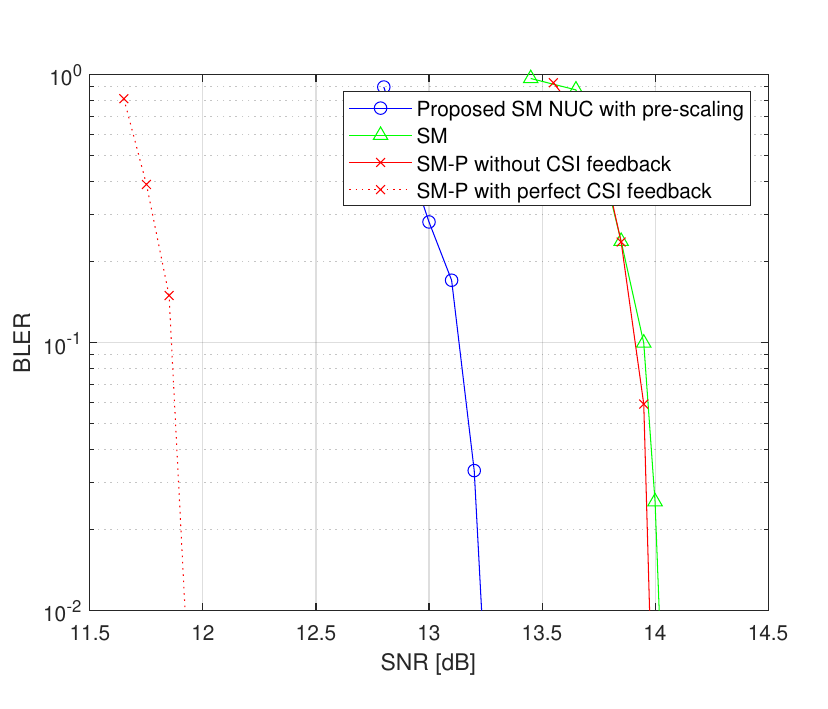}}
    \caption{The BLER result for $N_t=2$ and MCS 16.}
    \label{fig:2MCS-BLER}
\end{figure}

\begin{figure}[htbp]
    \centerline{\includegraphics[width=0.4\textwidth]{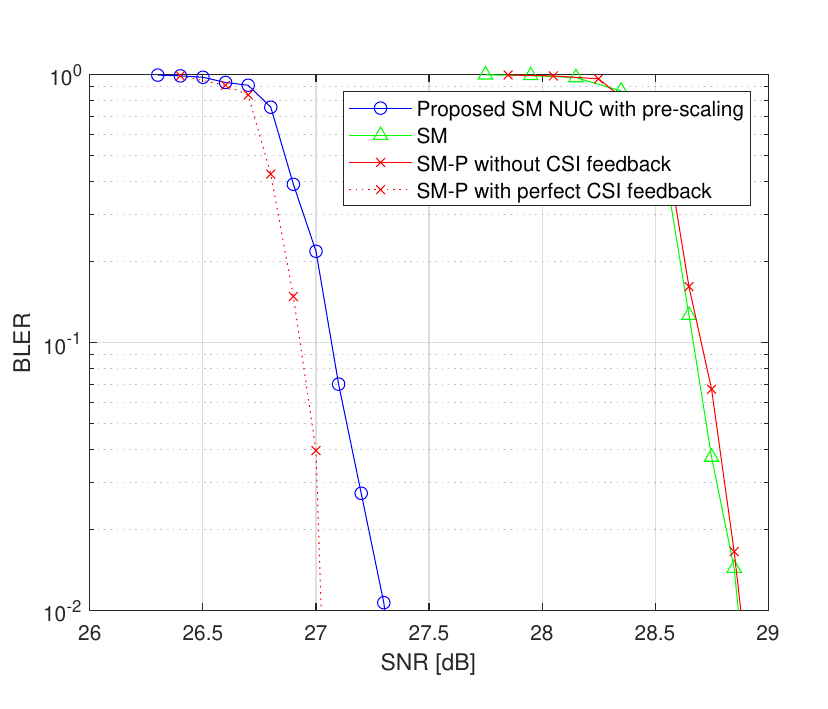}}
    \caption{The BLER result for $N_t=4$ and MCS 28.}
    \label{fig:4MCS-BLER}
\end{figure}

The obtained NUCs and pre-scaling coefficients are then used in the MISO system. Without CSI feedback, we compared the block error rate (BLER) performance of the proposed scheme, conventional SM, and SM-P in \cite{SM-P}, i.e., in \eqref{sm-p} $\alpha_k  =e^{j \theta_k}$ for all $k$. The BLER performance of SM-P with perfect CSI feedback is also tested. The criterion is to select the SNR that provides BLER of ${10}^{-2}$. For $N_t=2$ and MCS 16, which uses the NUC with $M=16$ and pre-scaling in \figref{fig:2MCS}(f), the BLER result is shown in \figref{fig:2MCS-BLER}. It is observed that SM-P with perfect CSI feedback outperforms SM by about 2.2 dB. But without CSI feedback, the performance of SM-P is very close to SM. On the contrary, the proposed scheme can bring a performance improvement for about 0.9 dB. For $N_t=4$ and MCS 28, which uses the NUC with $M=64$ and pre-scaling in \figref{fig:2MCS}(i), the BLER result is shown in \figref{fig:4MCS-BLER}. It is also observed that the performance of SM-P is very close to SM without CSI feedback, but the proposed scheme outperforms SM by about 1.6 dB.

\begin{table*}[htbp]
\renewcommand\arraystretch{1.5}
    \centering
    \caption{BLER preformance for $N_t=2$.}
    \begin{threeparttable}
    \begin{tabular}{|c|c|c||c|c|c|c|c|c|}
    \hline
        \rule{0pt}{12pt}
        MCS & M & \makecell{Code Rate \\$R$} & \makecell{Code Rate\tnote{1} \\$R'$} & SM [dB] & SM-P [dB] & \makecell{SM-P without \\feedback [dB]} & \makecell{Proposed NUC with \\pre-scaling [dB]} & \textbf{Gain}\tnote{2} [dB] \\ \hline\hline
        0  & 4  & 0.1190 & 0.0793 & -1.3 & -3.6 & -1.3 & -1.5 & \textbf{0.2} \\ \hline
        4  & 4  & 0.3088 & 0.2058 & 2.4 & 0.2 & 2.4 & 2.1 & \textbf{0.3} \\ \hline
        9  & 4  & 0.6564 & 0.4376 & 6.6 & 4.6 & 6.7 & 6.4 & \textbf{0.2} \\ \hline
        10 & 16 & 0.3287 & 0.2626 & 7.8 & 5.2 & 7.9 & 7.6 & \textbf{0.2} \\ \hline
        13 & 16 & 0.4760 & 0.3807 & 10.6 & 8.0 & 10.6 & 10.3 & \textbf{0.3} \\ \hline
        16 & 16 & 0.6405 & 0.5124 & 14.1 & 11.9 & 14.0 & 13.2 & \textbf{0.9} \\ \hline
        17 & 64 & 0.4271 & 0.3660 & 15.8 & 12.6 & 15.8 & 14.7 & \textbf{1.1} \\ \hline
        22 & 64 & 0.6459 & 0.5536 & 20.2 & 17.4 & 20.2 & 19.3 & \textbf{0.9} \\ \hline
        28 & 64 & 0.9189 & 0.7877 & 26.4 & 23.7 & 26.4 & 25.4 & \textbf{1.0} \\ \hline
    \end{tabular}
    \begin{tablenotes}
        \footnotesize
        \item[1] To be consistent with the BICM efficiency in the standard, $R'\times \log_{2}{MN_t}=R\times \log_{2}{M}$.
        \item[2] Gain is the performance gain of proposed NUC with pre-scaling,  calculated by $SNR_{SM}-SNR_{proposed}$.
      \end{tablenotes}
    \end{threeparttable}
    \label{tb:2}
\end{table*}
\begin{table*}[htbp]
\renewcommand\arraystretch{1.5}
    \centering
    \caption{BLER preformance for $N_t=4$.}
    \begin{threeparttable}
    \begin{tabular}{|c|c|c||c|c|c|c|c|c|}
    \hline
        \rule{0pt}{12pt}
        MCS & M & \makecell{Code Rate \\$R$} & \makecell{Code Rate\tnote{1} \\$R'$} & SM [dB] & SM-P [dB] & \makecell{SM-P without \\feedback [dB]} & \makecell{Proposed NUC with \\pre-scaling [dB]} & \textbf{Gain}\tnote{2} [dB] \\ \hline\hline
        0  & 4  & 0.1190 & 0.0595 & 1.7 & -2.9 & 1.7 & 0.8 & \textbf{0.9} \\ \hline
        4  & 4  & 0.3088 & 0.1544 & 5.2 & 1.4 & 5.2 & 4.3 & \textbf{0.9} \\ \hline
        9  & 4  & 0.6564 & 0.3282 & 8.8 & 6.3 & 8.7 & 8.4 & \textbf{0.4} \\ \hline
        10 & 16 & 0.3287 & 0.2191 & 12.4 & 10.1 & 12.4 & 10.8 & \textbf{1.6} \\ \hline
        13 & 16 & 0.4760 & 0.3173 & 14.0 & 9.3 & 14.0 & 12.8 & \textbf{1.2} \\ \hline
        16 & 16 & 0.6405 & 0.4270 & 16.9 & 15.0 & 16.9 & 15.6 & \textbf{1.3} \\ \hline
        17 & 64 & 0.4271 & 0.3203 & 20.3 & 16.2 & 20.3 & 18.2 & \textbf{2.1} \\ \hline
        22 & 64 & 0.6459 & 0.4844 & 23.9 & 20.9 & 23.8 & 22.1 & \textbf{1.8} \\ \hline
        28 & 64 & 0.9189 & 0.6892 & 28.9 & 27.1 & 28.9 & 27.3 & \textbf{1.6} \\ \hline
    \end{tabular}
    \begin{tablenotes}
        \footnotesize
        \item[1] To be consistent with the BICM efficiency in the standard, $R'\times \log_{2}{MN_t}=R\times \log_{2}{M}$.
        \item[2] Gain is the performance gain of proposed NUC with pre-scaling,  calculated by $SNR_{SM}-SNR_{proposed}$.
      \end{tablenotes}
    \end{threeparttable}
    \label{tb:4}
\end{table*}

More BLER performance results with with differect MCSs for $N_t=2$ and $N_t=4$ are shown in Table \ref{tb:2} and Table \ref{tb:4}, respectively. It can be seen that, without CSI feedback, the performance of SM-P is almost identical to SM. On the contrary, the proposed NUC with pre-scaling scheme can provide a gain up to 1.1 dB and 2.1 dB for $N_t=2$ and $N_t=4$, respectively. Hence, the proposed scheme is an effective and feasible alternative when CSI feedback is infeasible. Moreover, it can also be observed that, for MCSs with larger modulation order $M$, the proposed scheme can provide higher gain, and for the same MCSs, the proposed scheme can always provide higher gain at $N_t=4$ than at $N_t=2$. This is because higher modulation order and more TAs allow for more freedom in optimization. Therefore, we conclude that the proposed scheme can provide higher gain for the case with larger $M$ and $N_t$.
\addtolength{\topmargin}{0.04in}
\section{Conclusion}  \label{v}
This paper proposes a capacity-based joint optimization design scheme of NUC and pre-scaling, for situations where CSI feedback is unavailable. In MISO system with Rayleigh channel, the scheme has been shown to provide up to 1.1 dB and 2.1 dB error performance gain for $N_t=2$ and $N_t=4$, respectively. Besides, the proposed scheme can provide higher gain with larger modulation order and the number of TAs. Thus, the proposed scheme is a promising technology for the absence of CSI feedback. It will be interesting to research the solution to reduce the complexity of calculating BICM capacity in the future. Our future work will also involve the application of the proposed scheme to scenarios with inaccurate CSI estimation and more advanced SM techniques.

\section*{Acknowledgment}
This paper is supported in part by National Natural Science Foundation of China Program(62271316, 62101322), the Fundamental Research Funds for the Central Universities and Shanghai Key Laboratory of Digital Media Processing (STCSM 18DZ2270700).  The corresponding author is Yin Xu (e-mail: xuyin@sjtu.edu.cn).

\bibliographystyle{IEEEtran}
\bibliography{IEEEabrv,mybib}



\end{document}